\documentclass[epj]{svjour}
\usepackage{graphics}\usepackage{amsmath}
\usepackage{labelfig}

\usepackage{mdwlist}

\newcommand{\tr}{\mathrm{tr}\, }

\title{Individual and collective dynamics of self-propelled soft particles}

\author{M. Tarama\inst{1}, Y. Itino\inst{1}, A. M. Menzel\inst{2,1} \and T. Ohta\inst{1}}
\institute{
  \inst{1} Department of Physics, Kyoto University, Kyoto, 606-8502, Japan \\
  \inst{2} Institut f\"ur Theoretische Physik II: Weiche Materie, Heinrich-Heine-Universit\"at D\"usseldorf, Universit\"atsstra{\ss}e 1, D-40225 D\"usseldorf, Germany }

\date{Received: date / Revised version: date}

\PACS{
{05.45.-a}{Nonlinear dynamics and chaos} \and
{47.54.-r}{Pattern selection; pattern formation}
}

\abstract{
Deformable self-propelled particles provide us with one of the most important nonlinear dissipative systems, which are related,  for example,  to the motion of microorganisms. It is emphasized that this is a subject of localized objects in non-equilibrium open systems.  We introduce a coupled set of ordinary differential equations to study various dynamics of individual soft particles due to the nonlinear couplings between migration, spinning and deformation. By introducing interactions among the particles, the collective dynamics and its collapse are also investigated by changing the particle density and the interaction strength. 
We stress that assemblies of self-propelled particles also exhibit a variety of non-equilibrium localized patterns. 
}

\begin{document}

\maketitle

\section{Introduction}\label{sec:intro}
Existence of a localized object is one of the characteristic features in non-variational and non-equilibrium systems since  such a solution is
unstable in a variational system  in two and three dimensions, which has a Lyapunov functional or a free energy functional with a short range interaction in thermal equilibrium \cite{Derrick}.  Stable localized domains exist, for example, in excitable reaction-diffusion media which is inherently non-variational \cite{Ohta89}.   The  complex Ginzburg-Landau (CGL) equation with both an ordinary cubic nonlinearity and a cubic-quintic nonlinearity  also exhibits stable localized solutions \cite{Bekki,Brand94,Cross}. This complex Ginzburg-Landau equation is a reduced equation for oscillatory systems and nonlinear optics. For certain parameters, these localized solutions in the reaction-diffusion equations and the CGL equations  undergo translational motion 
which is called a drift bifurcation
\cite{Mikhailov,Purwins,Ohta,Nishiura}. 

Self-propulsion of an isolated domain or a droplet is common in many non-equilibrium systems. An oil droplet is an example if there are chemical reactions at its interface and the products affect the interfacial tension locally as a Marangoni effect  \cite{Nagai,Toyota1,Ban,Thutupalli,Kitahata}. Colloidal particles whose surfaces are decorated also exhibit translational motions under certain non-equilibrium conditions \cite{Kapral1,Kapral2,Sano,Squires,Showalter,Ebbens}.

Migration of living micro-organisms is also a typical phenomenon of self-propelled motion \cite{Keren,Bosgraaf,Li,Maeda,Sawai}. There are elaborated theories for specific living cells and bacterias \cite{Wada,Ishikawa,Sasai,Levine,Ziebert} as well as of artificial models  \cite{Kruse,Fu,Yeomans}.

It is emphasized that all of the above systems except for hard colloids are deformable and therefore, nonlinear couplings between shape deformation and translational motion play an important role in their dynamics.  In fact, all of the theories for biological objects  \cite{Wada,Ishikawa,Sasai,Levine} have considered the shape deformations. The biological motions are actually a deformation-induced migration.  Chemically reacting domains are also generally deformed depending on the
magnitude 
of the translational velocity \cite{Mikhailov,Nagai}, which is called a migration-induced deformation. As mentioned above, there are several theories that take shape deformations into consideration. However, those are developed to represent individual specific systems. 

We formulate the dynamics of deformable self-propelled objects from a unified view point since this is one of the fundamental nonlinear dynamics far from equilibrium. 
This prototype model approach follows the same spirit as the CGL equation which is a basic equation for oscillatory media independent of the details of the systems considered. 
The set of time-evolution equations for a deformable self-propelled particle has been derived by a symmetry argument near the bifurcation threshold where a motionless state becomes unstable and a translational motion appears \cite{OhtaOhkuma}.  One of the advantageous properties of the theory is that it is applicable both to two and three dimensions in a unified manner. The coefficients of the time-evolution equations which are left as unknown parameters in the phenomenological theory have been determined  in excitable reaction-diffusion equations  \cite{OOS,SHO}. Numerical and analytical results have been obtained \cite{Hiraiwa1,Hiraiwa2}.

In summary, there exist localized objects that are self-propelled and deformable. As examples, we mentioned liquid droplets driven by spatially inhomogeneous interfacial tension and soft migrating biological cells. Although these objects are complex and consist of many interacting building blocks, they can be viewed as one entity and treated in analogy to a single particle. Apart from that, spatially localized structures can arise from the interaction between many individual deformable self-propelled objects. Obvious examples in nature are bird and fish swarms, bacterial clusters, or herds of cattle. To understand the properties of these spatially localized superstructures, the collective behavior of self-propelled particles must be investigated. We briefly summarize recent approaches introduced for this purpose. After this, we concentrate on the behavior of single deformable self-propelled particles and aspects of their collective behavior. 

First, we shall describe the individual dynamics of deformable self-propelled particles that possess a spinning degree of freedom \cite{Tarama1,Tarama2}.  This is also motivated by the fact  that a spinning motion is observed for many  micro-organisms. {\it Listeria} is an example, which causes migration by actin polymerization and undergoes a helical motion  \cite{Listeria2,Listeria1,Crenshaw,Listeria3}. Flagellated bacteria such as {\it Escherichia coli} also exhibits 
a spinning motion by rotating helical filaments \cite{DiLuzio,Goldstein}. Theoretical studies have also been performed \cite{Goldstein,Chen}. 
A  spontaneous formation of spiral waves in a {\it Dictyostelium} cell has been discovered, which causes a spinning motion as well as migration and deformation \cite{Sawai}. We introduce a spinning variable that is coupled to the center-of-mass motion and the shape deformations. Although we have studied spinning dynamics both in two and three dimensions, we here mainly show the results in two dimensions. It is mentioned here that a similar model but without deformations has been investigated recently \cite{Loewen12}.

Collective dynamics of self-propelled objects is also an important subject in nonlinear science and non-equilibrium physics. In fact, there are a myriad  of studies of the motions of assemblies of interacting self-propelled particles. See refs. \cite{Vicsek2,Ramaswamy} and the earlier papers cited therein. Again, there are no intensive studies of deformable interacting particles. Here we review the  results of this subject  in two spatial dimensions \cite{Itino1,Itino2,Menzel}.  The interaction between a pair of particles is given by a Gaussian form with respect to the distance. We consider two cases. 
 In the first case (model I) the magnitude of the interaction depends on the relative angle of the elongation of the pair of particles such that a parallel configuration is more favorable \cite{Itino1,Itino2}. In the second case (model II) such an
 alignment 
  mechanism is not explicitly contained but the force from other particles affects the shape deformation directly \cite{Menzel}. We show the results of numeral simulations of these two models. 
Hydrodynamic effects possibly mediated by surrounding media are not considered.

The organization of this paper is as follows. First, in section \ref{sec:recent}, we briefly repeat recent model descriptions that show the emergence of localized structures from the collective behavior of mainly self-propelled 
point-like
particles. In section \ref{sec:model}, we introduce the model equations for a single deformable self-propelled particle with a spinning degree of freedom. Numerical results together with some analytical results are given in section \ref{sec:numerical_results_1}.   The models of interacting particles are introduced in section \ref{sec:model2}. Numerical simulations are described  in section \ref{sec:numerical_results}. Summary and discussion are given in section \ref{sec:discussion}.

\section{
Localized Structures found in Recent Models of Self-Propelled Particles}\label{sec:recent}

In this article we concentrate on polar self-propelled particles. This means that the particles cannot reverse their propulsion direction. Nematic particles, such as rigid granular rods in a vertically vibrated horizontal layer \cite{Narayan}, which can reverse their propulsion direction \cite{Chate06}, will not be covered in the following. 

The most famous particle-based model to study the onset of collective behavior in self-propelled particle crowds is probably the one introduced by Vicsek et al.\ \cite{Vicsek95}. Particles are treated as point-like and all of them are assigned the same magnitude $v_0$ of the self-propulsion velocity at each time step. This simplification leaves as the only degrees of freedom for each particle its position $\mathbf{r}^{(i)}$ and the orientation $\mathbf{\hat{v}}^{(i)}$ of its velocity vector $\mathbf{{v}}^{(i)}$, where the superscript $i$ labels the particles. The particle position is updated after each time step $\Delta t$ through a simple streaming 
\begin{equation}\label{vicsek_r}
\mathbf{r}^{(i)}(t+\Delta t) = \mathbf{r}^{(i)}(t) + v_0\, \mathbf{\hat{v}}^{(i)}(t)\Delta t. 
\end{equation}
In two spatial dimensions, the velocity orientation $\mathbf{\hat{v}}^{(i)}$ can be parameterized by one angle $\vartheta^{(i)}$. The only interaction between different particles consists of an alignment rule for their velocity vectors applied at each timestep: 
\begin{equation}\label{vicsek_theta}
\vartheta^{(i)}(t+\Delta t) = \langle \vartheta^{(j)}(t)\rangle_r + \eta^{(i)}(t). 
\end{equation}
Here, the angular brackets denote an average over the angular orientations of all particles $j$ within a spherical environment of radius $r$ centered at the location of particle $i$. This alignment rule promotes polar alignment of the particle velocity vectors. $\eta^{(i)}$ denotes an angular noise that counteracts the velocity alignment. 

The competition between the velocity alignment and the orientational noise leads us to the central point of the model. For high noise strengths (or low particle densities) the particle motion is disordered. When the noise strength is reduced (or the particle density is increased), a transition occurs to a state of orientationally ordered particle velocity vectors. This is a state of collective motion, since the particles coherently migrate into one direction.

\begin{figure}[t]
  \begin{center}
  \resizebox{0.37\textwidth}{!}{
    \includegraphics{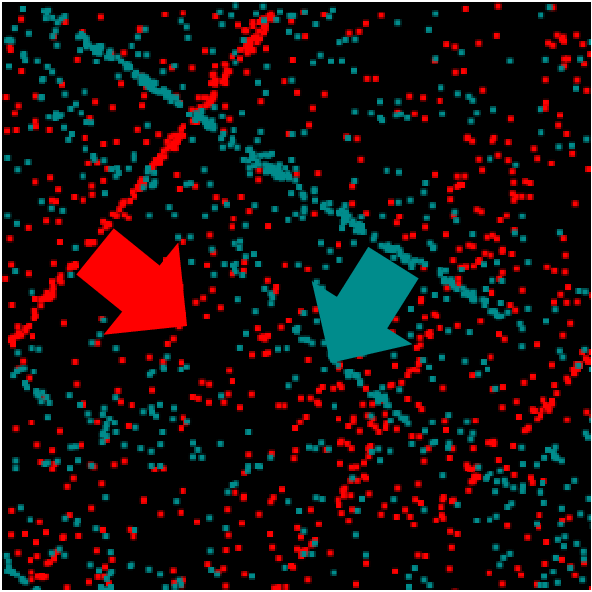}
  }
  \caption{
    (Color online) Traveling density bands in a model system of the Vicsek type as obtained from an initially disordered state. At the investigated particle densities, high-density bands form and propagate collectively perpendicularly to their elongation direction. In the low-density background region, particles propel in an uncorrelated disordered way. Periodic boundary conditions are applied. In this figure, particles of two separate species are shown, indicated by the two different colors. This figure is partially reproduced from ref.~\cite{Menzel12}. 
  } \label{fig:travelingband}
  \end{center}
\end{figure}

The transition scenario has been thoroughly investigated for particles following a polar alignment rule as described above \cite{Gregoire04,Raynaud08,Gregoire08,Peruani11,Farrell12}.
In the topical context of localized objects and textures, the most interesting scenarios occur around the transition corresponding to the onset of collective motion. There the particle density becomes highly spatially inhomogeneous and traveling localized structures appear. Typically they have the form of traveling density bands \cite{Gregoire08}. Within each band, the density is high and the particle velocities collectively order generally perpendicular to the band elongation in this case. Outside the band region, the particle density is low and disordered motion prevails. A similar band structure has been observed experimentally \cite{Joanny,Schaller}.
A numerical example of a traveling band structure is depicted in Fig.~\ref{fig:travelingband} for particles of two different 
kinds of species, 
each of which 
organized in a traveling density band \cite{Menzel12}. 
The density bands corresponding to the two different species have different orientations of their collective self-propulsion velocities.

The transition to collective motion has also been investigated for
particles conforming to nematic velocity alignment rules (i.e.\ antiparallel or polar alignment of single velocity vectors are equally preferred) \cite{Raynaud08,Peruani11,Ginelli10,Peruani10}. In the latter case, the migration direction was found along a density band. Although several aspects of the transition depend on fine details of the numerical implementation \cite{Gregoire08,Albano}, the appearance of spatially localized objects seems to be rather robust. Excluded volume effects can supplement or replace the alignment rules \cite{Peruani06,Baskaran08,Boehme11,Hagan12,Wensink12}. When the self-propulsion velocity changes with the density, further types of localized textures are found \cite{Farrell12}. We will present a new, qualitatively different propagating band structure in section \ref{sec:numerical_results}. 

Apart from these particle based approaches, self-propel-led particle crowds were investigated by macroscopic ``hydro\-dynamic-like'' continuum equations \cite{Toner95,Toner98,Ramaswamy02,Ramaswamy05,Marchetti10}. The basic variables are the coarse-grained density field $\rho(\mathbf{r},t)$ and a coarse-grained velocity field $\mathbf{v}(\mathbf{r},t)$ \cite{Toner98}. Often, the latter is replaced by a polar vector order parameter field $\mathbf{P}(\mathbf{r},t)$ assuming a typical magnitude $v_0$ of the self-propulsion velocity. It is crucial to note that the model lacks Galilean invariance if only the motion of the self-propelled particles is included, without taking into account the surrounding medium, substrate, etc. The corresponding symmetry-allowed equations of motion can be written in the form \cite{Toner98}
\begin{equation}
\partial_t \rho + \nabla\cdot(\rho\mathbf{v}) = 0, 
\end{equation}
\begin{eqnarray}
\lefteqn{\partial_t \mathbf{v} + \lambda_1(\mathbf{v}\cdot\nabla)\mathbf{v} 
  + \lambda_2(\nabla\cdot\mathbf{v})\mathbf{v} + \lambda_3\nabla\mathbf{v}^2} \nonumber\\
& = & {} \alpha\mathbf{v} -\beta\mathbf{v}^2\mathbf{v} -\nabla P(\rho) 
      +D_B\nabla(\nabla\cdot\mathbf{v}) \nonumber\\
&& {}+ D_T\nabla^2\mathbf{v}+D_2(\mathbf{v}\cdot\nabla)^2\mathbf{v} +\mathbf{f}. 
\end{eqnarray}
Here, the scalar pressure term $P(\rho)$ is a function of the density, $\mathbf{f}$ is a noise term, and $\lambda_1$, $\lambda_2$, $\lambda_3$, $\alpha$, $\beta$, $D_B$, $D_T$, and $D_2$ are phenomenological parameters. For $\alpha>0$ correlated collective motion appears. 

\begin{figure}[t]
  \begin{center}
  \resizebox{0.48\textwidth}{!}{
    \includegraphics{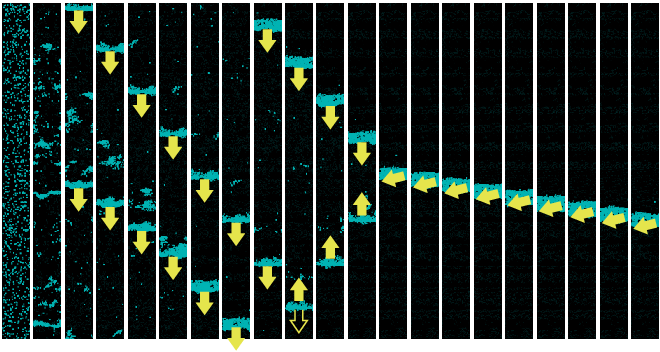}
  }
  \caption{
    (Color online) Collision of two propagating density bands in a particle model of the Vicsek type. 
    Shown is a time series of snapshots from left to right, with equal time intervals between successive snapshots. From the initially disordered state two traveling density bands form. First, the two bands travel approximately downward. After a while, we invert the velocity directions of all particles in the bottom 
band, so that the collective propulsion velocity of the whole band is reversed and it now travels upward. When the two bands collide, one compound band forms with a collective propagation orientation nearly perpendicular to the long side of the calculation box. Interpenetration and passing of the two colliding bands, resulting again in two propagating density bands after the collision, is not observed here. 
  } \label{fig:collision}
  \end{center}
\end{figure}

It was demonstrated that the continuum models contain propagating front solutions \cite{Marchetti10,Bertin09}. These were related to the traveling density band structures described above for the particle-based models \cite{Bertin09}. Often in this context, the propagating density bands are referred to as solitary wave structures both in the particle and in the continuum models \cite{Gregoire04,Gregoire08,Bertin09,Gopinath}. Still the analysis of the stability of these structures and the stronger connection between the particle and continuum picture require further investigation. The latter concerns, for example, the shape of the bands that were observed to have a steep front and a more smoothly decaying rear part in the particle simulations \cite{Bertin09}. 
It seems to us that an interpenetration and passing of these structures
is at most rarely observed for the traveling density bands in the particle picture. To address this point, we performed a simple numerical collision experiment of two propagating density bands. A time series of snapshots corresponding to this collision scenario is illustrated in Fig.~\ref{fig:collision}. For this purpose, we numerically solved a particle model of the Vicsek type [see Eqs.~(\ref{vicsek_r}) and (\ref{vicsek_theta})] in a simulation box  of high aspect ratio and periodic boundary conditions. Under the given conditions, two density bands form from the initially disordered state and propagate approximately along the long side of the numerical box into the same direction. To make these two bands collide, we reverse the velocity orientations of all particles within one of the two density bands. In this way, the collective migration direction of the band is inverted. The two bands now head for collision. After colliding, the two bands form one compound object. 
Again, this point requires further investigation.

Finally, density fluctuations that are significantly larger than in equilibrium systems were predicted to occur in self-propelled particle systems \cite{Toner98,Ramaswamy02,Ramaswamy05}. We consider the number fluctuations $\delta N$ in a subregion of fixed volume. Then the standard deviation of the number fluctuations is expressed as $\sqrt{\langle\delta N\rangle^2}\propto N^{\zeta}$, where $N$ is the averaged particle number in the subregion. For equilibrium systems $\zeta=0.5$. In active polar systems in two spatial dimensions, an exponent of up to $\zeta=1$ was predicted \cite{Toner98,Ramaswamy02,Ramaswamy05}. These large number fluctuations have been measured numerically. They were related to the formation of immobile clusters in systems of self-propelled particles that interact via short-ranged soft repulsive forces \cite{Fily,Redner}. Such clusters form another interesting example of localized states in self-propelled particle systems.

\section{Model Equations for Isolated  Particles}\label{sec:model}

The time-evolution of deformable self-propelled particles  in two dimensions is represented in terms of the center-of-mass velocity 
 ${\bf v}$ and the symmetric tensor $\tens{S}$ characterizing deformations. The components of this tensor are given by 
 \begin{equation}
S_{\alpha \beta} = s \left( n_{\alpha} n_{\beta}-\frac{1}{2}\delta_{\alpha \beta} \right), 
\end{equation}
where  the  unit vector ${\bf n}$ is parallel to the long axis of a deformed elliptical particle and $s>0$ is the degree of deformation from a circular shape. The spinning motion can be incorporated by an antisymmetric tensor 
 $\tens{\Omega}$ which takes the form 
\begin{equation}
\tens{\Omega}  
 = 
 \left[
 \begin{array}{cc}
 0 & \omega \\
 -\omega & 0
 \end{array}
 \right]
 \label{eq:1.5}
\end{equation}
with $\omega$ a dynamical variable. The components of $\tens{\Omega}$ will be denoted as $\Omega_{\alpha\beta}$.

By considering possible couplings among these variables and retaining some relevant nonlinear terms, the set of time-evolution equations is given by \cite{Tarama1}
\begin{eqnarray}
\frac{d v_{\alpha}}{dt} 
 &=& \gamma v_{\alpha} -v_{\nu}v_{\nu} v_{\alpha} -a_1 S_{\alpha \beta}v_{\beta} -a_2 \Omega_{\alpha \beta} v_{\beta} ,
 \label{eq:1.1}\\
\frac{d S_{\alpha\beta}}{dt}
 &=& -\kappa S_{\alpha\beta} +b_1 ( v_{\alpha} v_{\beta} -\frac{v_{\nu}v_{\nu}}{2} \delta_{\alpha\beta}) \nonumber\\
 &&+ b_2 ( S_{\alpha\nu}\Omega_{\nu\beta} -\Omega_{\alpha\nu}S_{\nu\beta}) + b_3 \Omega_{\alpha\nu} S_{\nu\lambda} \Omega_{\lambda\beta} ,
 \label{eq:1.2}\\
\frac{d \Omega_{\alpha\beta}}{d t} 
 &=& -\frac{\partial G}{\partial \Omega_{\alpha\beta}} 
 + 4c_2 S_{\alpha\nu} \Omega_{\nu\lambda} S_{\lambda\beta} ,
 \label{eq:1.3}
\end{eqnarray}
where 
\begin{equation}
G \equiv \frac{\zeta}{2} \tr {\tens{\Omega}}^2 +\frac{1}{4} \tr {\tens{\Omega}}^4  ,
 \label{eq:1.4}
\end{equation}
and $a_1$, $a_2$, $b_1$, $b_2$, $b_3$ and $c_2$ are the coupling constants. 
 On the right hand side of Eq.~(\ref{eq:1.3}), there could be a second-order coupling term $c_1 ( S_{\alpha\nu}\Omega_{\nu\beta} +\Omega_{\alpha\nu}S_{\nu\beta})$, which vanishes identically  in two dimensions. 
The repeated indices imply summation. The coefficient $\gamma$ takes either negative or positive values. 
Throughout the present paper, the parameters  $\kappa$ and $\zeta$ are assumed to be positive.

Equations~(\ref{eq:1.1}) and  (\ref{eq:1.2}) in the absence of  $\tens{\Omega}$  exhibit a bifurcation at $\gamma=\gamma_c$ defined by \cite{OhtaOhkuma}.
\begin{equation}
\gamma_c \equiv \frac{\kappa^2}{a_1 b_1} +\frac{\kappa}{2} = \frac{\kappa (1+B)}{2B}, 
 \label{eq:1.7} 
\end{equation}
where
\begin{equation}
 B= \frac{a_1b_1}{2\kappa}. 
 \label{eq:B}
\end{equation}
A particle undergoes a  straight motion  for $0 < \gamma < \gamma_c$ whereas it becomes unstable for $\gamma>\gamma_c$ and the particle exhibits a circular motion in which the trajectory of the center of mass displays a closed circle.  
There are two cases of straight motion depending on the sign of $b_1$. If $b_1$ is positive, the particle elongates along the direction of the migration velocity whereas, if it is negative, the elongation is perpendicular to the velocity. 

It is remarked here that, when $a_1b_1<0$ and $a_2=b_2=b_3=0$, eqs. (\ref{eq:1.1}) and (\ref{eq:1.2}) are variational and can be written as 
\begin{eqnarray}
\frac{dv_\alpha}{dt}&=&-L_1\frac{\delta F}{\delta v_\alpha} \\
\frac{dS_{\alpha\beta}}{dt}&=&-L_2\frac{\delta F}{\delta S_{\alpha\beta}}
 \label{eq:vS}
\end{eqnarray}
with positive constants $L_1$ and $L_2$. The Lyapunov functional $F$ takes the form 
\begin{eqnarray}
F&=&\int d\vec{r}\Big[\frac{\alpha_1}{2}v_{\nu}v_{\nu}+\frac{\alpha_2}{4}(v_{\nu}v_{\nu})^2 \nonumber \\
&+&\frac{\alpha_3}{2}S_{\nu\lambda}S_{\nu\lambda}+\alpha_4 v_{\nu}S_{\nu\lambda}v_{\lambda}\Big]
\end{eqnarray}
where $\alpha_i$ ($i=1,..,4$) are constants which should satisfy the necessary conditions to guarantee the concaveness of $F$.  A term
$(1/2)L_2\alpha_4 v_{\nu}v_{\nu} \delta_{\alpha\beta}$ must further be added to eq.~(\ref{eq:vS}) in order to obtain eq.~(\ref{eq:1.2}) and to ensure that 
$S_{\alpha\alpha}=0$.

\section{Numerical Results  of Spinning Motion}\label{sec:numerical_results_1}

\begin{figure}[t]
  \begin{center}
  \resizebox{0.4\textwidth}{!}{
  \includegraphics{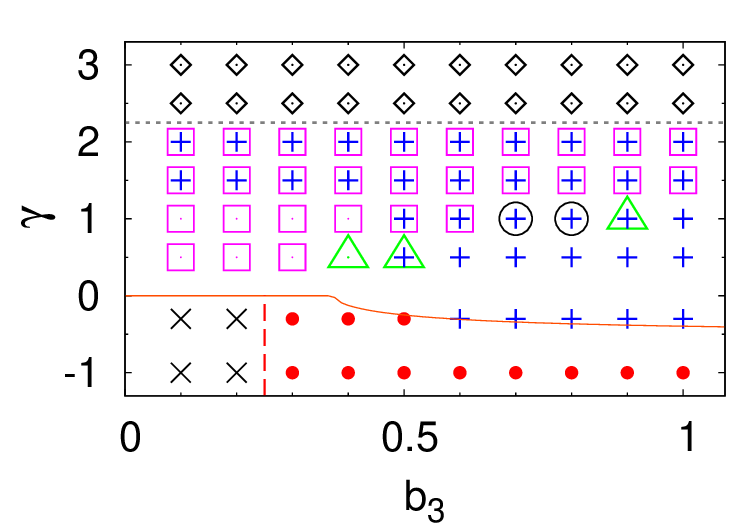}
}
      \caption{(Color online) 
      Dynamical phase diagram obtained by solving Eqs.~(\ref{eq:1.1})--(\ref{eq:1.4}) numerically. 
      The  meaning of the symbols is given in the text.
          The bifurcation boundary from the motionless state to the spinning state without migration is indicated by the broken line,   the thin solid line for the boundary between the $v=0$ and the $v\ne0$ region and
      the dotted line for the threshold at which  the state $\omega=0$ becomes unstable are obtained by a theoretical analysis \cite{Tarama1}.
      The parameters are $\kappa=0.5$, $a_1=-1$, $a_2=0.75$, $b_1=-0.5$, $b_2=1$, $\zeta=2$, and $c_2=2$. 
The bifurcation threshold $\gamma_c$ defined by eq. (\ref{eq:1.7}) is given by $\gamma_c=0.75$. This figure is reproduced from ref.  \cite{Tarama1}.
      } \label{fig:diagram}
  \end{center}
\end{figure}

In this section, we show the results of numerical simulations of the model equations (\ref{eq:1.1})--(\ref{eq:1.4}) in two dimensions  \cite{Tarama1}.
The coefficients are set to  $\kappa=0.5$, $a_1=-1$, $a_2=0.75$, $b_1=-0.5$, $b_2=1$, $\zeta=2$, and $c_2=2$, whereas the coefficients $\gamma$ and $b_3$ are varied.
Figure~\ref{fig:diagram} shows the dynamical phase diagram obtained by  numerical simulations. A fourth order Runge-Kutta method with a time increment of $\delta t = 10^{-4}$ was used.

\begin{figure}[t]
  \begin{center}
  \resizebox{0.4\textwidth}{!}{
  \includegraphics{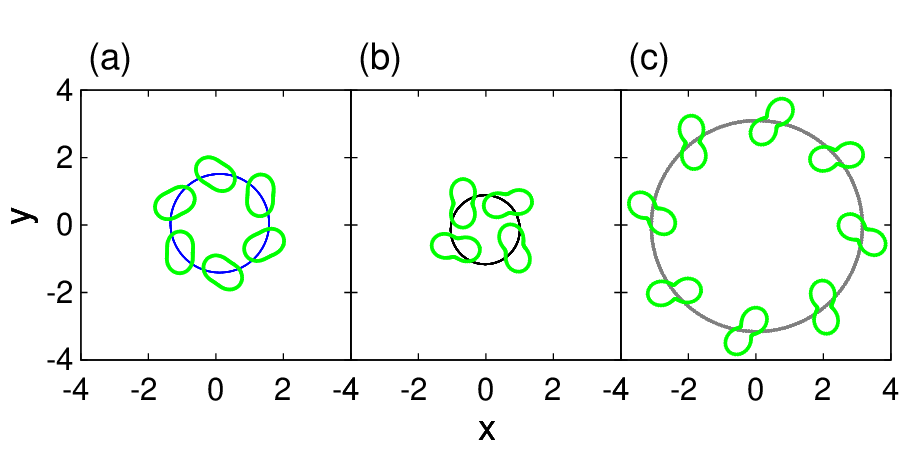}
}
      \caption{(Color online) 
      Rotation and orbital revolution of a particle in the counter-clockwise direction. A trajectory in real space is displayed for ({a})  revolution I motion for $\gamma=1$ and $b_3=0.5$,  ({b}) revolution II motion for  $\gamma=1$ and $b_3=0.9$,  and ({c}) circular motion  for $\gamma=3$ and $b_3=0.5$.  The particle size in ({a})--({c})  is reduced by a factor of $1/4$ for the sake of clarity. This figure is reproduced from ref.  \cite{Tarama1}.
      } \label{fig:rotating}
  \end{center}
\end{figure}

In the region indicated by the cross and the filled-circle in Fig.~\ref{fig:diagram}, the center of mass of a particle is motionless, i.e. $v=0$, and the value of the internal rotation $\omega$ takes a finite constant value. 
In the cross region, the shape of the particle is circular without deformation, $s=0$, and hence a spinning motion is irrelevant in spite of a finite $\omega$. 
We call this state a motionless state. 
On the other hand, a particle in the region of the filled-circle is elliptically elongated, i.e. the magnitude of the deformation $s$ is a finite constant, with the angle of the elongation direction varying monotonically in time. 
Therefore, this motion is called a spinning motion.

There also exist several dynamical states where the center of mass of a particle undergoes orbital motions. 
The square symbols and the plus symbols in Fig.~\ref{fig:diagram} represent two different circular motions, which we call the revolution I and revolution II states, respectively. 
In both of these revolution states, all of the variables $\omega$, $s$, and $v$ take finite constant values. The angle between the velocity vector and the direction of elongation is constant in time. 
The trajectories of the revolution I and II motions in real space are displayed in Figs.~\ref{fig:rotating}({a}) and ({b}), where some snapshots of a particle that undergoes counter-clockwise rotation are shown. 
It is noted in Fig.~\ref{fig:diagram} that there is a region where the revolution I and  revolution II states coexist. 
There is another orbital revolution which appears in the region indicated by the diamonds in Fig.~\ref{fig:diagram}.
We call this state a circular state since $\omega =0$ while the values of $v$ and $s$ are finite constants and the angle of the propagation direction $\phi$ and the elongation direction $\theta$ vary monotonically but their difference $\psi=\theta-\phi$ is constant. The trajectory and the particle shape of a counter-clockwise circular motion in real space are displayed in Fig.~\ref{fig:rotating}({c}).  
Since an internal rotation does not occur, $\omega=0$, this is simply the circular motion obtained in Ref.~\cite{OhtaOhkuma}.

\begin{figure}[t]
  \begin{center}
  \resizebox{0.4\textwidth}{!}{
  \includegraphics{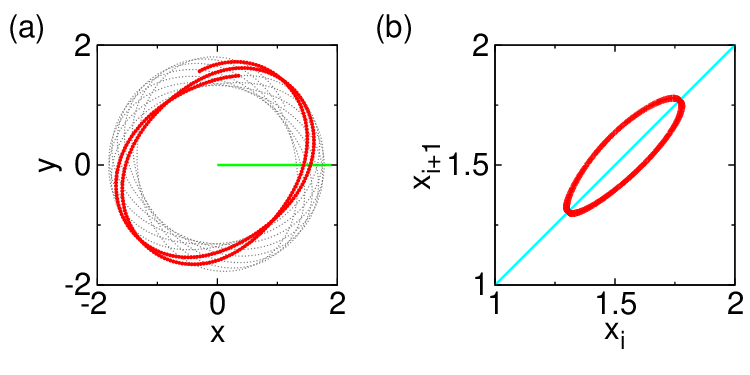}
}
      \caption{(Color online) 
     ({a}) Trajectory and ({b}) return map of the quasi-periodic I motion for $\gamma=1$ and $b_3=0.8$  in the real space.
     The trajectory for a shorter  time interval is indicated by the thick solid line in ({a}).
           The return map  is obtained at the Poincar\'{e} section indicated  by the horizontal line in ({a}).  This figure is reproduced from ref.  \cite{Tarama1}.
     } \label{fig:qp1}
  \end{center}
\end{figure}

\begin{figure}[t]
  \begin{center}
  \resizebox{0.4\textwidth}{!}{
  \includegraphics{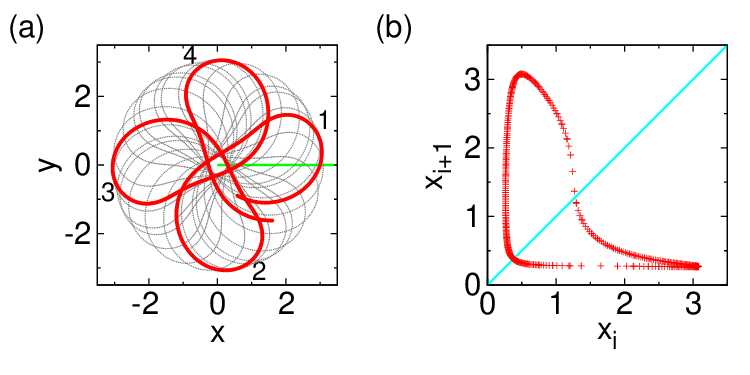}
  }
      \caption{(Color online) 
      ({a}) Trajectory and ({b}) return map of the quasi-periodic II motion for $\gamma=1$ and $b_3=0.9$  in the real space.
           The trajectory for a shorter time interval is shown by the thick solid line in ({a}) together with the chronological order  indicated by the numbers 1, 2, 3,  and 4. 
            The return map is obtained at the Poincar\'{e} section indicated by the horizontal line in ({a}).  This figure is reproduced from ref.  \cite{Tarama1}.
      } \label{fig:qp2}
  \end{center}
\end{figure}

Apart from these three orbital revolutions, there are two types of quasi-periodic motions. 
In the region of the unfilled circles around $\gamma=1$ and $b_3=0.75$  in Fig.~\ref{fig:diagram}, a quasi-periodic I state appears, whose trajectory of a counter-clockwise rotation is displayed in Fig.~\ref{fig:qp1}({a}). 
In Fig.~\ref{fig:qp1}({b}), we show the return map of this motion, which is obtained at the Poincar\'{e} section indicated by the horizontal line in Fig.~\ref{fig:qp1}({a}). 
In this state, all of the variables $v$, $\phi$, $s$, $\theta$, and $\omega$, as well as the difference $\psi$ are time-dependent. 
There is another quasi-periodic motion in the region of the triangles in Fig.~\ref{fig:diagram}, which is called a quasi-periodic II state. One example of the trajectory in real space is shown in Fig.~\ref{fig:qp2}({a}) for $\gamma=1$ and $b_3=0.9$, where a particle rotates in the counter-clockwise direction. Some parts of the trajectory are highlighted by the thick solid line together with the number indicating the chronological order. 
In this state, all of the variables $\omega$, $v$ and $s$ are non-zero, and both of $\phi$ and $\theta$ as well as  their difference $\psi$ are time-dependent. 
In Fig.~\ref{fig:qp2}({b}), we show the return map of the quasi-periodic II motion, which is obtained at the Poincar\'{e} section at the horizontal line in Fig.~\ref{fig:qp2}({a}). 
Note that both the quasi-periodic  I and II states coexist with the revolution II state (plus symbols) in Fig.~\ref{fig:diagram}.

\begin{figure}[t]
  \begin{center}
  \resizebox{0.4\textwidth}{!}{
  \includegraphics{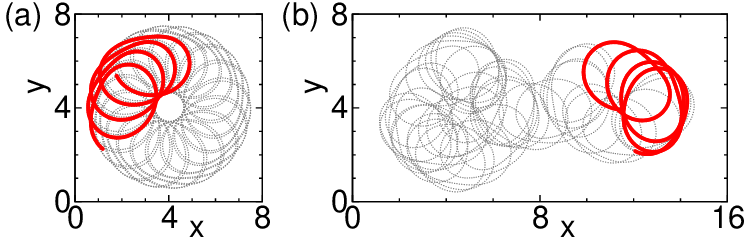}
}
      \caption{(Color online) 
      Trajectory in the real space of ({a}) the period-doubling state and ({b}) the chaotic state. 
      The trajectory for  a shorter  time interval is shown by the thick solid line.
      The parameters are set to $\gamma=1$ as well as ({a}) $b_3=0.82$ and ({b}) $b_3=0.844$.  This figure is reproduced from ref.  \cite{Tarama1}.
      } \label{fig:trace_pd}
  \end{center}
\end{figure}

By increasing the values of $b_3$, the quasi-periodic I state becomes unstable and a period-doubling bifurcation occurs. The corresponding trajectory in the real space is displayed in Fig.~\ref{fig:trace_pd}({a}).
Further increase of $b_3$ leads to a chaotic motion as shown in Fig.~\ref{fig:trace_pd}({b}).
This chaotic state eventually disappears for larger values of $b_3$ and, in turn,  the quasi-periodic II state appears.
This series of the dynamical transitions from the quasi-periodic I state  for $b_3=0.8$ to the quasi-periodic II state  for $b_3=0.9$ are not shown  in Fig.~\ref{fig:diagram} due to the size of the grid.

In order to study  some of the bifurcations analytically, we made a reduction of the variables. The degree of deformation $s$ and the relative angle $\psi$ were retained \cite{Tarama1}. Through a linear stability analysis based on the simplified set of equations, the bifurcation lines indicated in Fig.~\ref{fig:diagram}  were obtained.

\section{Models for Collective Dynamics} \label{sec:model2}

There are few theoretical studies of the collective motion of self-propelled particles that include deformations. Therefore, here, we introduce two kinds of model systems of interacting self-propelled particles. 
The time-evolution of the {\it i}-th particle obeys a set of equations for the position of the center of mass  $\vec{r}^{(i)}$, the velocity  $\vec{v}^{(i)}$ and the deformation tensor $\tens{S}^{(i)}$. The spinning degree of freedom is not considered in the study of collective motion. 

Then, model I takes the following equations \cite{Itino1,Itino2}
\begin{eqnarray}
 \frac{d}{dt}r^{(i)}_{\alpha} &=& v^{(i)}_{\alpha} ,
 \label{r} \\
 \frac{d}{dt}v^{(i)}_{\alpha} &=& \gamma v^{(i)}_{\alpha} -(\vec{v}^{(i)})^2 v^{(i)}_{\alpha} - aS^{(i)}_{\alpha\beta}v^{(i)}_{\beta} + f^{(i)}_{\alpha} +\xi^{(i)}_{\alpha},
 \label{v} \\
 \frac{d}{dt}S^{(i)}_{\alpha\beta} &=& -\kappa S^{(i)}_{\alpha\beta} + b\left(v^{(i)}_{\alpha}v^{(i)}_{\beta}-\frac{1}{2}v^{(i)}_{\nu}v^{(i)}_{\nu}\delta_{\alpha\beta}\right) ,
 \label{S}
\end{eqnarray}
where $\gamma$, $\kappa$, $a$ and $b$ are positive constants. 
The deformation  tensor for the {\it i}-th particle is defined by
 \begin{equation}
S^{(i)}_{\alpha \beta} = s^{(i)} \left( n^{(i)}_{\alpha} n^{(i)}_{\beta}-\frac{1}{2}\delta_{\alpha \beta} \right).
\end{equation}
The interaction between a pair of two particles is given by
the term $ {\bf f}^{(i)} $ in eq. (\ref{v}), which  is the force acting on the $i$-th particle and takes the form 
\begin{equation}
{ \bf f}^{(i)} = K\sum\limits_{j=1}^N{\bf F}_{ij}Q_{ij} .
 \label{f}
\end{equation}
Here, the index $i$ is not summed over, 
$K$ is a positive constant, $N$ is the total number of particles, and 
\begin{equation}
{ \bf F}_{ij}= -\frac{\partial U_1({\bf r}_{ij})}{\partial{\bf r}_{ij}} ,
   \label{F}
\end{equation}
with ${\bf r}_{ij}={\bf r}^{(i)}-{\bf r}^{(j)}$.
The potential $ U_1({\bf r}_{ij})$ is assumed to be given by 
\begin{eqnarray}
U_1({\bf r}_{ij})=\exp\Big[-\frac{{\bf r}_{ij}^2}{2\sigma^2}\Big],
\label{U1}
\end{eqnarray}
with $\sigma=1$ throughout the present paper.
The other factor $Q_{ij}$ in eq. (\ref{f}) represents an alignment mechanism of the particles and is defined by
\begin{eqnarray}
 Q_{ij}= 1+\frac{Q}{2}\mathrm{tr}\left\{(\tens{S}^{(i)}-\tens{S}^{(j)})^{2}\right\} ,
\label{Q}
\end{eqnarray}
where $Q$ is a positive constant. 
In two dimensions, we have
\begin{eqnarray}
 Q_{ij} 
&=&1+\frac{Q}{4}\Big[(s^{(i)}-s^{(j)})^2 \nonumber \\
&+& 4s^{(i)}s^{(j)}\sin^2(\theta^{(i)} - \theta^{(j)})\Big]  
\label{Q2}
\end{eqnarray}
from the parameterization ${\bf n}^{(i)}=(\cos\theta^{(i)}, \sin\theta^{(i)})$.
This indicates that when the elongated directions of two particles are parallel to each other, the repulsive interaction is weaker than in a perpendicular configuration. 
The last term  $\xi^{(i)}_{\alpha}$ in eq. (\ref{v})  is a Gaussian noise satisfying
\begin{eqnarray}
 \langle \xi^{(i)}_{\alpha}(t)\rangle &=& 0 ,\\
 \langle \xi^{(i)}_{\alpha}(t)\xi^{(j)}_{\beta}(s)\rangle&=&\eta^2\delta_{ij}\delta_{\alpha\beta}\delta(t-s).
\end{eqnarray}
Since self-propulsion is a phenomenon far from equilibrium, we do not impose the fluctuation-dissipation relation.
For simplicity, we do  not add a noise term in eq. (\ref{S}) for the deformation tensor.

The model II takes the following time-evolution equations \cite{Menzel}
\begin{eqnarray}
 \frac{d}{dt}v^{(i)}_{\alpha} &=& \gamma v^{(i)}_{\alpha} -v^{(i)}_{\nu}v^{(i)}_{\nu}v^{(i)}_{\alpha} - aS^{(i)}_{\alpha\beta}v^{(i)}_{\beta} + \mu g^{(i)}_{\alpha} , \label{v2} \\
 \frac{d}{dt}S^{(i)}_{\alpha\beta} &=& -\kappa S^{(i)}_{\alpha\beta} + b'\left(v^{(i)}_{\alpha}v^{(i)}_{\beta}-\frac{1}{2}v^{(i)}_{\nu}v^{(i)}_{\nu}\delta_{\alpha\beta}\right)  \nonumber \\
&- &c'\left(g^{(i)}_{\alpha}g^{(i)}_{\beta}-\frac{1}{2}g^{(i)}_{\nu}g^{(i)}_{\nu}\delta_{\alpha\beta}\right)  ,
 \label{S2}
\end{eqnarray}
with an additional equation for the position that is the same as eq. (\ref{r}) and $b'=b w$ as well as $c'=cw$ with $w=1-2S^{(i)}_{\alpha\beta}S^{(i)}_{\alpha\beta}$. The multiplicative factor $w$ is introduced to prevent unphysically large deformations 
in the situation of a very close contact of a pair of particles. 
 Such a difficulty does not occur actually for the parameters investigated in model I. 
As in model I, we choose $\gamma>0$ and $\kappa>0$ to enforce self-propulsion and make a circular particle stable when it is motionless. We set $a<0$ and $b>0$ so that the dynamics of a single particle is variational as mentioned at the end of section  \ref{sec:model}. The velocity tends to orient parallel to the long axis of the elliptic deformation for each particle. An isolated  particle moves ballistically on straight trajectories both in model I and II. 

The force $\mathbf{g}^{(i)}$ mimics soft steric interactions between the particles, which is given by $g_{\alpha}^{(i)}= - \partial U_2/\partial r_{\alpha}^{(i)}$. The soft interaction potential $U_2$ is defined by \cite{Menzel} 
\begin{equation}\label{U2}
U_2 = \exp\left[-\frac{1}{2\sigma^2}\mathbf{r}_{ij}\cdot(\tens{I}-\tens{S}^{(i)})\cdot(\tens{I}-\tens{S}^{(j)})\cdot\mathbf{r}_{ij}\right],
\end{equation}
where $\tens{I}$ is the unit tensor and the deformed shape of a particle has been taken into account.

\section{Numerical Results of Interacting Particles}\label{sec:numerical_results}

First, we show the numeral results for model I. The parameters are fixed as $a=1$, $b=0.5$, $\gamma=\kappa=\sigma=1$, $K=5$ and $Q=50$. 
The density is varied by changing the system size.
In order to check the size effect of  the ordered collective dynamics, several values of the  total number of particles  are  chosen:  $N= 512$, $2048$, $8192$, $32768$. We introduce the order parameter as
\begin{equation}\label{OP}
  \Phi = \Big|\frac{1}{N}\sum\limits_{j=1}^{N}e^{2i\theta^{(j)}}\Big|,
\end{equation}
where $\theta^{(j)}$ denotes the angle of the elongated direction of the $j$-th particle.
This represents the degree of directional ordering of elongated particles and corresponds to the nematic order parameter in the case of liquid crystals. 
We may define another order parameter in terms of the direction of the migration velocity as
\begin{equation}\label{OP2}
\Phi_{v}=\Big|\frac{1}{N}\sum\limits_{j=1}^{N}e^{i\phi^{(j)}}\Big|
\end{equation}
with $\phi^{(j)}$  characterizing the propagating direction.
This is  a polar order parameter. We have checked numerically that  there is no essential difference between the values of the two order parameters in the ordered state obtained in model I.

\begin{figure*}[t]
  \begin{center}
	\resizebox{0.9\textwidth}{!}{
	  \includegraphics{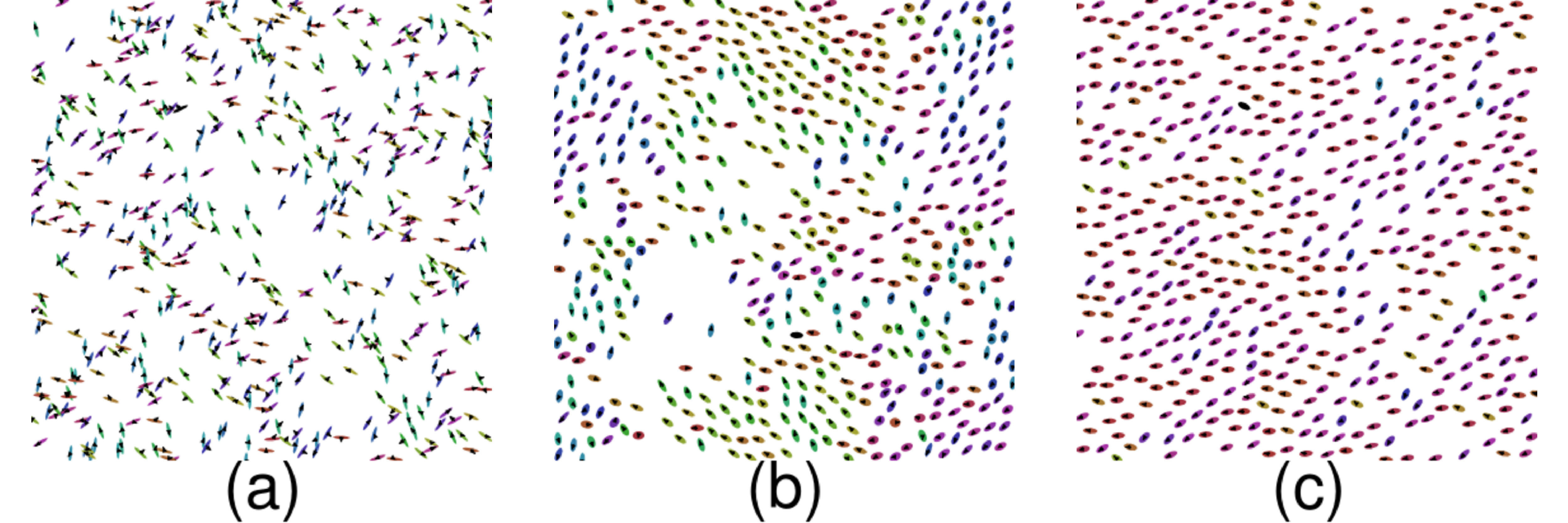}
}
    \caption{(Color online)
   Snapshots of the ordering process at ({a}) $t=0$, ({b}) $t=1260$ and ({c}) $t=1792$ in model I for $\rho=0.0625$, $\eta=0.6$ and $N=512$.  
   The colors indicate the direction of elongation of each particle.  This figure is reproduced from ref. \cite{Itino2}.
     }
      \label{fig:original_traces_2}
  \end{center}
\end{figure*}

\begin{figure*}[t]
  \begin{center}
	\resizebox{0.95\textwidth}{!}{
	  \includegraphics{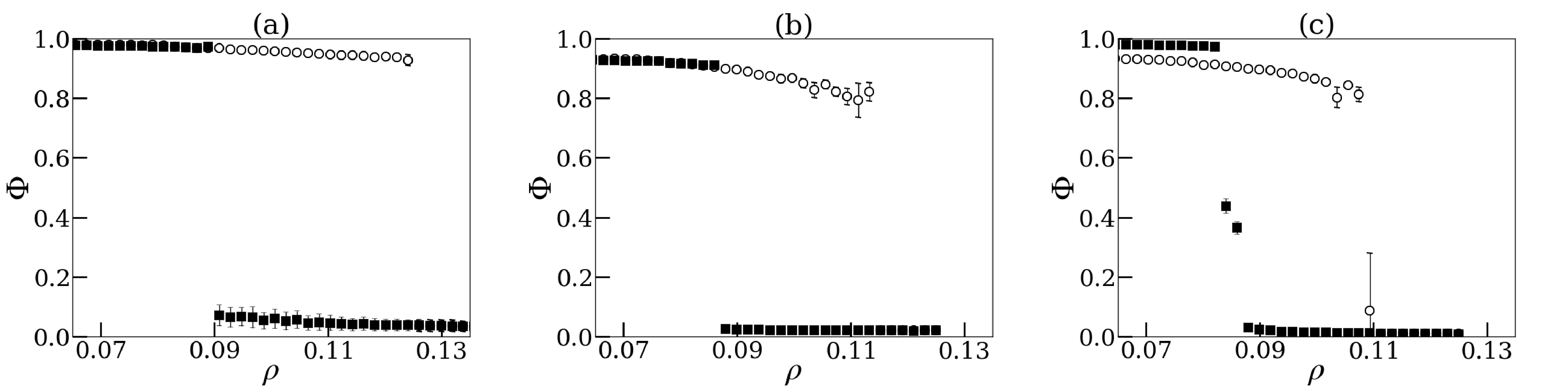}
}
    \caption{Hysteresis loop in model I for ({a}) $N=512$, ({b}) $N=2048$, and ({c}) $N=8192$ as a
function of $\rho$ for  $\eta=0.1$. 
The data indicated by the unfilled circles are obtained by increasing the density whereas the black squares are the data obtained by decreasing the density. The data points in ({c}) are explained further  in the text.
The error bar indicates  the  standard deviation of $\Phi$ around the average
obtained in the time interval  $T=4096$.   This figure is reproduced from ref. \cite{Itino2} 
with a change of data symbols.  } \label{fig:hysteresis}
  \end{center}
\end{figure*}

Figure~\ref{fig:original_traces_2} displays the ordering process of the self propelled particles starting from a disordered state ({a}) \cite{Itino1}. As time elapses, through the intermediate state ({b}), the particles tend to migrate to the same direction on average as shown in ({c}).  It is noted, however, that the positional order is not complete in ({c}). Periodic boundary conditions are imposed in all the figures.

One of the characteristic features of model I (and model II also as shown below) is that the ordered state becomes unstable for sufficiently large values of the density. This can be seen in Fig.~\ref{fig:hysteresis} \cite{Itino2}. The density is progressively changed by an increment $\Delta \rho=1/1024$ and the system is relaxed for the duration time $T=4096$  after each change in density.  Although  a small $N$-dependence is observed, it is clear that a transition from the ordered state to the disordered state exists and there is a hysteresis. We have checked that the kinetic energy of each particle is almost constant independent of the density. Therefore, a possibility of raising ``temperature" by compressing the system is excluded.  We have evaluated numerically the self-intermediate scattering function and the mean square displacement of a test particle in the disordered state and have confirmed that the disordered state is neither a glass state nor a jamming state \cite{Itino1}.

\begin{figure}[t]
	\begin{center}
	\resizebox{0.45\textwidth}{!}{
	  \includegraphics{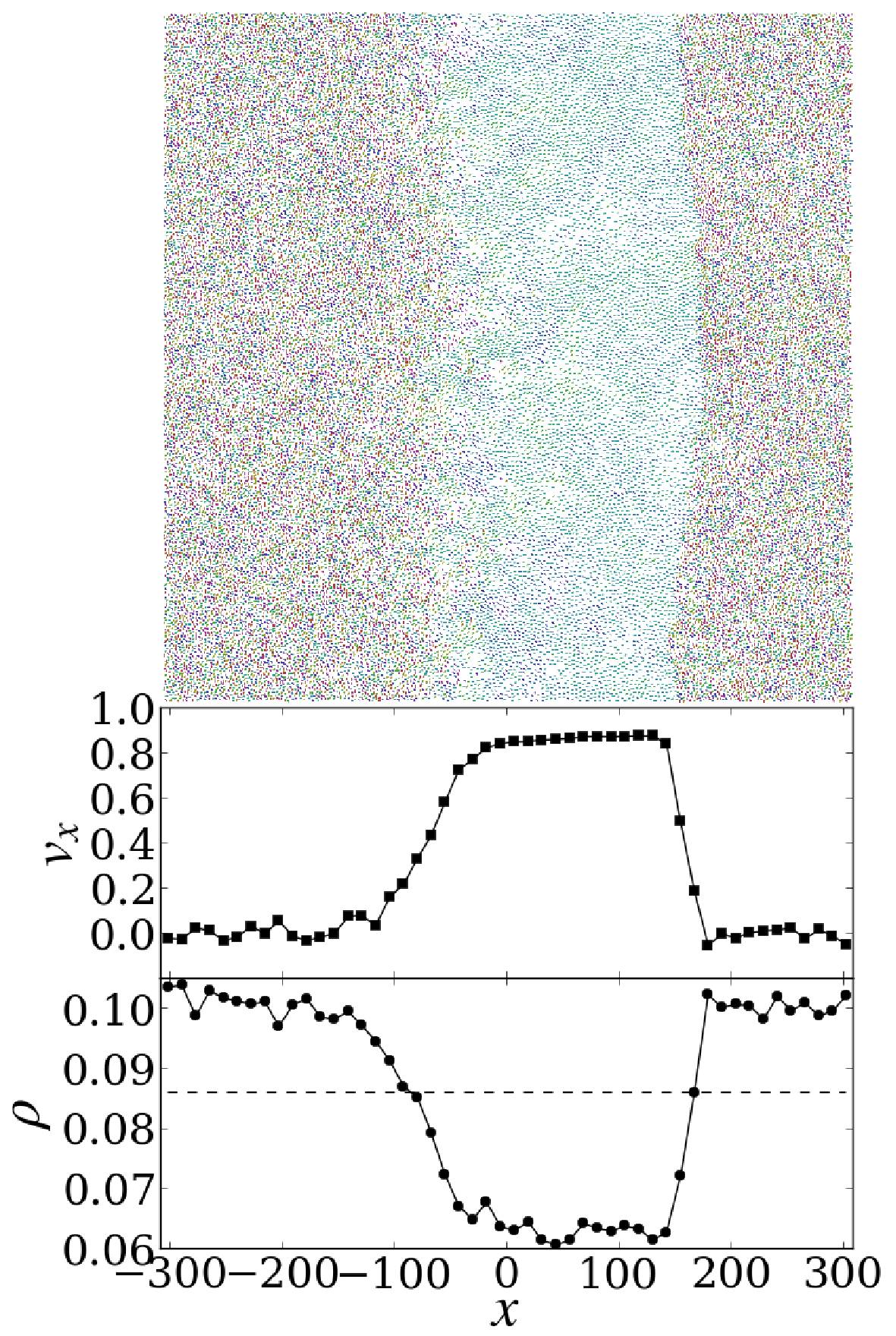}
	}
\caption{(Color online)
      Propagating band structure of the ordered state (lighter) and the disordered state (darker) in model I for $N=32768$, $\rho=0.086$ and $\eta=0.05$. 
      The $x$-components of the velocity and the density averaged over the $y$ direction are also shown as a function of the $x$ axis. The average of the order
      parameter is about $\Phi=0.28$.
          }
           \label{fig:jammed}
	\end{center}
\end{figure}

One unexpected phenomenon occurs in Fig.\ref{fig:hysteresis}({c}) for $N=8192$. By decreasing the density, the system does not jump from the disordered state to the ordered state directly but it is trapped at an intermediate state with $\Phi \approx 0.3$--$0.4$. In the trapped phase, there is a dynamical coexistence of the ordered state and the disordered state. Figure~\ref{fig:jammed} indicates the snapshot of the trapped phase where we have shown the results for  $N=32768$ and the noise intensity $\eta=0.05$ since the phenomenon is observed  more frequently  for larger system size and smaller noise intensity.  
In ref. \cite{Itino2}, the dynamical coexistence was obtained for $N=8192$ but no clear band structures were found as shown in Fig.~\ref{fig:jammed}.
The ordered state is in the lighter vertical region with low density whereas the disordered state is in the darker regions with higher density. It is noted that each particle is propagating  to the right in the ordered state while the boundaries between the ordered and the disordered states are moving to the left. A similar band structure has been observed in ref. \cite{Gregoire08} where, however, the density of the ordered state is higher than the density of the disordered state and furthermore the moving direction of the boundary between the ordered and the disordered states is the same as that of individual particles in the ordered state. The present dynamics shown in Fig.~\ref{fig:jammed}  is analogous (though not strictly in one dimension) to the motion of a boundary between jammed and unjammed regions observed on freeways.  By a slight further decrease of the density, the system enters into the ordered state with  $\Phi \approx 1$ indicated by the black squares. This value of $\Phi$ is larger than the value at the branch of the increasing density $\Phi \approx 0.95$ shown by the unfilled circles. This might come from the fact  that the ordered state in the branch of the increasing density was generated starting from a complete random initial condition whereas the branch with $\Phi \approx 1$ was caused from the trapped phase where 
 the system is partially ordered.

\begin{figure}
  \begin{center}
\resizebox{0.45\textwidth}{!}{
	  \includegraphics{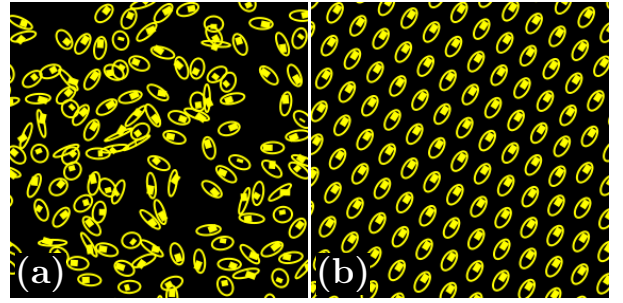}
}
    \caption{(Color online)
    Snapshots from the ordering process of a representative sample, starting from a disordered state of random initial conditions in model II. The sample is still in the disordered state in ({a}) at $t=600$, but has orientationally and translationally ordered in ({b}) at $t=12500$. System parameters: $N=300$, $\rho=0.14$, $\mu=0.01$, $c=2.5$.  This figure is reproduced from ref. \cite{Menzel}. 
    }\label{fig_orderedstate}
  \end{center}
\end{figure}

\begin{figure}
  \begin{center}
	\resizebox{0.45\textwidth}{!}{
	  \includegraphics{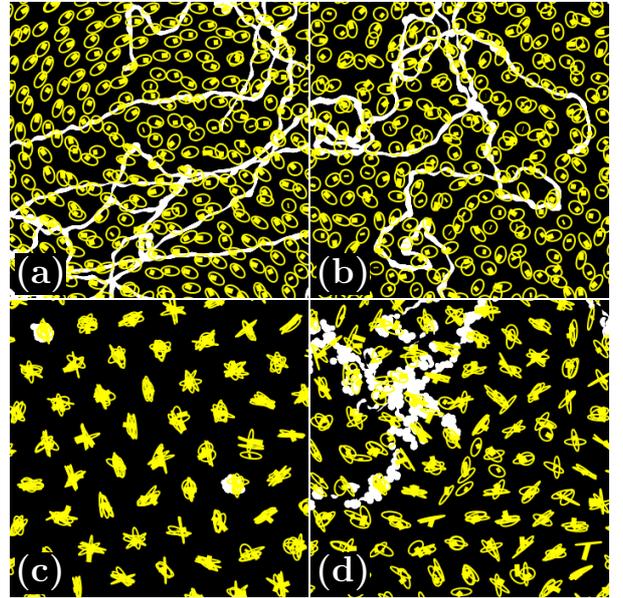}
	}
    \caption{(Color online)
     Illustration of the collective behavior at high densities after sufficiently slow compression from a hexagonally ordered polar state in model II. The initial density was $\rho=0.2$ in all cases. Trajectories of selected particles are indicated in white (line thickness decreases with increasing velocity). Different panels correspond to different cut-off radii $r_c$ and densities: ({a}) $r_c=5\sigma$, $\rho=0.84$, ({b}) $r_c=5\sigma$, $\rho=0.93$,  ({c}) $r_c=2\sigma$, $\rho=1.27$, ({d}) $r_c=1.5\sigma$, $\rho=1.27$. Other system parameters: $N=1000$, $\mu=0.05$, $c=2.5$.  This figure is partially reproduced from ref. \cite{Menzel}. }
     \label{fig_breakdown}
  \end{center}
\end{figure}

Now we present numerical results of model II given by  eqs.\ (\ref{r}), (\ref{v2}) and (\ref{S2}) \cite{Menzel}. We set $\gamma=0.5$, $\kappa=\sigma=1$, $a=-1$, and $b=0.5$ unless stated otherwise.  Figure~\ref{fig_orderedstate}({a}) shows an initially disordered state for $\rho=0.14$ and  $\mu=0.01$ after starting from random initial conditions. Ellipsoidal shapes highlight the particle deformations and the thick bars within the ellipsoids indicate the velocity orientations. As time proceeds, an ordering process occurs. The final ordered state is depicted in Fig.~\ref{fig_orderedstate}({b}), where the particles constitute a hexagonal lattice. In addition, the velocity and deformation directions collectively order so that the depicted particles in the figure are uniformly propagating to the upper-right direction. 

In model I, there is a transition from the ordered state to the disordered state by increasing the density. We have also examined this scenario in model II by compressing the system sufficiently slowly. 
We have used a truncated (and shifted) interparticle potential of the form of eq.~(\ref{U2}) with a cut-off distance $r_c$ \cite{Menzel}. The results for several choices of $r_c$ are shown in Fig.~\ref{fig_breakdown}. In this figure, example trajectories of single particles are indicated in white. The local line thickness decreases with increasing magnitude of the recorded velocity, and only part of the numerical calculation box is shown. At the initial density of $\rho=0.2$, all samples showed hexagonal polar order similar to Fig.~\ref{fig_orderedstate}({b}). Particle sizes are adjusted for visualization 
in Figs. ~\ref{fig_orderedstate} and \ref{fig_breakdown} below.

Figures.~\ref{fig_breakdown}({a}) and ({b}) correspond to a cut-off distance $r_c=5\sigma$. After global order has broken down, we still find some local order ({a}), which vanishes under further compression ({b}). The velocity magnitudes along a trajectory in ({b}) are highly inhomogeneous, indicating that collisions frequently occur. Therefore this disordered state is again neither a glass state nor a jamming state. 
Instead, it appears rather fluid-like. 
Slowly expanding the systems to their initial density, all of these structures turned back into the polar hexagonal crystal structure. We also observed hysteretic effects  \cite{Menzel2}.  
It is noted that the density (area fraction) of order $\rho\sim O(1)$ is relatively large. 

A possible explanation for the fluidification under compression is the Gaussian-core interaction potential applied between the particles. This potential determines the core-repulsion through the force term with the coefficient $\mu$ in Eq.~(\ref{v2}). In equilibrium systems, i.e.\ in the non-self-propelled case, the Gaussian-core potential is known to promote fluidification of the crystalline structure when the density is increased \cite{Stillinger,Prestipino1,Prestipino2,Prestipino3,Rex}. To test this assertion, we replaced the Gaussian-core potential by a truncated and shifted Lennard-Jones potential, the so-called Weeks-Chandler-Andersen potential, to calculate the repulsive force terms with the coefficient $\mu$. We still found a hysteretic breakdown of the orientational velocity ordering under compression. 
However, the hexagonal lattice structure was preserved in our numerical tests and we did not observe the fluidification as for the Gaussian-core repulsion \cite{Menzel2}. 

Interesting behavior occurs at smaller cut-off lengths. Figure~\ref{fig_breakdown}({c}) shows an example for $r_c=2\sigma$. The boundedness of the potential plays an important role. It allows putting several particles at the same spatial location with only finite energetic penalty. It is known that at high densities non-self-propelled particles with soft bounded core interactions  can form a cluster crystal in which several particles are stacked at each lattice site \cite{Likos,Schmidt}. The state in Fig.~\ref{fig_breakdown}({c}) is a self-propulsion version of a cluster crystal. The deformed particles are not ordered orientationally.  As indicated by the localized trajectories, the particles are trapped at the lattice points and the clusters are essentially immobile.  Although a lattice order exists, one may say that this is a kind of jamming. The situation may be connected to the effect that particles in a plane under compression stack on top of each other at high densities
 . Due to this effect, the area density $\rho$ can become larger than $1$. It is mentioned that, quite recently, interacting self-propelled ``rigid" particles with a soft repulsive potential have been investigated and a jamming state at high density has been found \cite{Henkes}.
If we further reduce the cut-off length to $r_c=1.5\sigma$, this structure is more mobile, as indicated by the trajectories in Fig.~\ref{fig_breakdown}({d}). In effect, the particles can move over the plane. We may call this scenario a ``cluster fluid-like'' state.

\begin{figure}[t]
  \begin{center}
	\resizebox{0.45\textwidth}{!}{
	  \includegraphics{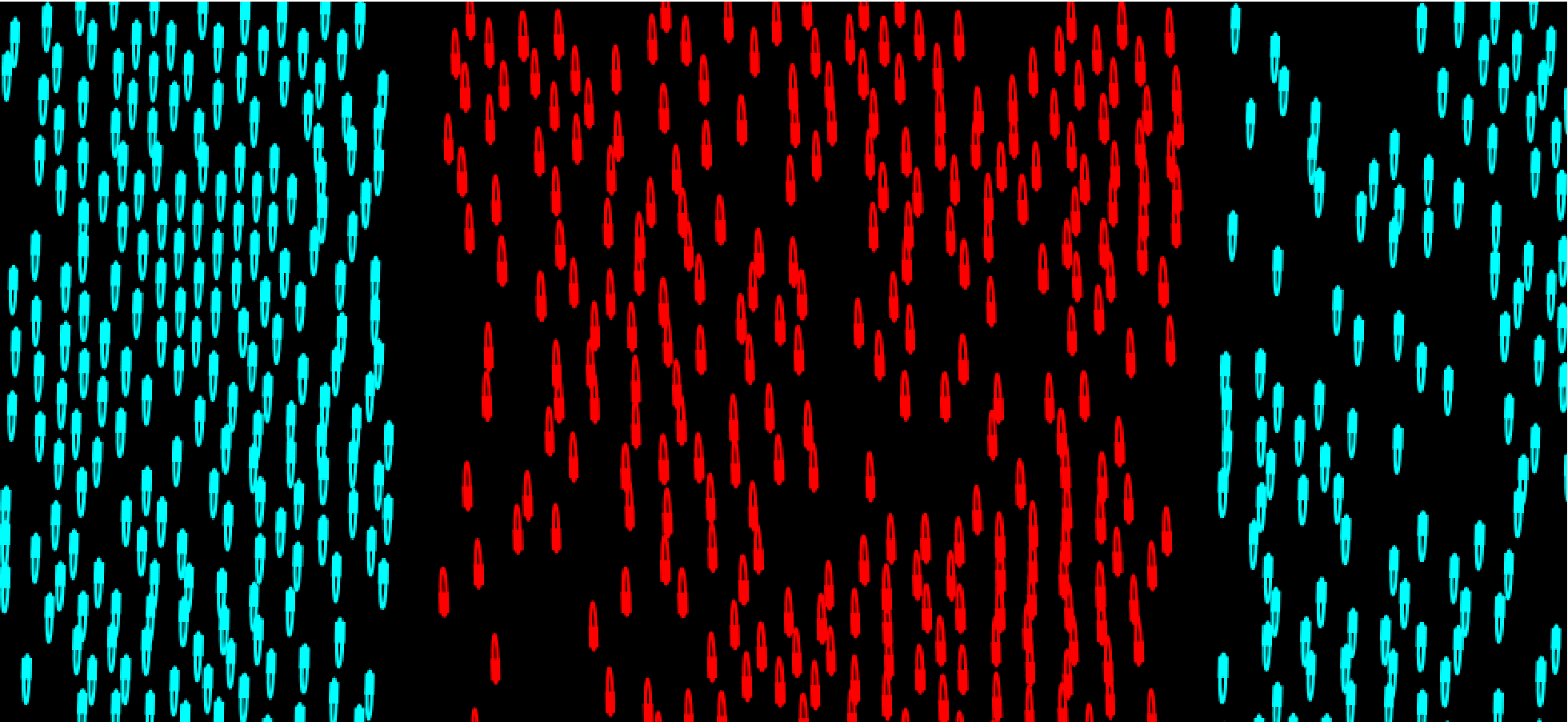}
	}
    \caption{(Color online)
   Laning in a system of soft deformable self-propelled particles of high aspect ratio in model II. Red particles move downward, turquoise ones upward. Only part of the numerical calculation box is shown. System parameters: $N=1200$, $\rho=0.14$, $a = -10$, $\mu=0.03$, $c = 1.2$. This figure is reproduced from ref. \cite{Menzel}.  }
\label{fig_laning}
  \end{center}
\end{figure}

Finally it is remarked that laning occurs in model II as displayed in Fig.~\ref{fig_laning}. Again we started from random initial conditions to obtain this result. The velocity alignment is nematic on a global scale and polar within each lane. Therefore the value of the order parameter $\Phi$ defined by eq.~(\ref{OP}) is finite whereas the global value of the order parameter $\Phi_v$ given by eq.~(\ref{OP2}) is small. 
Nonzero but not too high values of $c>0$ seem to support the macroscopic separation into different lanes. In the example of Fig.~\ref{fig_laning}, we measured a sample-averaged degree of deformation $\langle s \rangle\approx0.8$. This corresponds to an aspect ratio of about $2.33:1$.

\section{Discussion}\label{sec:discussion}

We have studied the dynamics of deformable self-propelled particles theoretically in two dimensions.  Both, individual and collective dynamics, have been considered. In the individual dynamics, the coupling of the spinning motion with the center-of-mass motion and the deformations has been introduced. Even for an isolated single particle, the dynamics is non-trivial exhibiting  several types of motion, such as quasi-periodic motions and chaos. 

In the collective dynamics, we have introduced two kinds of models with a Gaussian-core potential.  One of the common properties of both models is that the collective motion collapses  at high density, which occurs as a discontinuous transition with hysteresis.  It has been well known that a crystal of (non-self-propelled) particles with Gaussian-core potential undergoes melting by increasing the density \cite{Stillinger,Prestipino1,Prestipino2,Prestipino3,Rex}. This  behavior  is called reentrant fluidity in thermal equilibrium. It is highly likely that the disordering in the self-propelled particles at high density  is also due to the Gaussian-core potential. However, one has to distinguish two kinds of orders in the present non-equilibrium system, that is, orientational order of the propagating direction and the positional order. For example, Fig.~\ref{fig:original_traces_2}({b}) shows an orientational order with less complete positional order. This should be compared to F
 ig.~\ref{fig_orderedstate}({b}) where both orders are developed. There are also two cases in the disordered state at high density. One is the case where both orientational order and positional order are broken. The other is that orientational order does not exist but particles constitute a cluster crystal as in Fig.~\ref{fig_breakdown}({c}). Therefore, this question needs to be further investigated both, analytically and numerically,  to clarify the properties of the transition for non-equilibrium self-propelled particles.

A band structure as in Fig.~\ref{fig:jammed}  appears in model I in the vicinity of the threshold at high density where a disordered state becomes unstable. This formation seems to be possible only when the system size is sufficiently large and the density is changed extremely slowly. A similar phenomenon may also be found in model II.   Furthermore,  laning is obtained in model II as shown in Fig.~\ref{fig_laning} whereas it is not observed in model I. The pair-wise interaction (\ref{Q}) depends only on the relative elongation direction but not on the relative angle of the propagating velocities. Therefore, a possibility of laning is not excluded in model I. One of the reasons why it has not been obtained is that the aspect ratio in the simulations of model I is about 1.5, which is much smaller than the aspect ratio 2.33 in model II. That is, the particles in model II are slender compared to those in model I. 

It is emphasized that the present study indicates  that  formation of localized structures has a fascinating hierarchical structure. Our basic equations (\ref{eq:1.1}) and (\ref{eq:1.2}) for the velocity of the center of mass and the deformation tensors  without $\tens{\Omega}$ can be derived as a set of time-evolution equations for a single domain 
(a localized object)
in excitable reaction-diffusion equations \cite{OOS,SHO}. By introducing the interaction among the particles, we obtain other kinds of localized objects such as the band structure in Fig.~\ref{fig:jammed}   and the cluster lattice in Fig.~\ref{fig_breakdown}({c}). These findings imply that one needs to develop  theoretical and experimental studies further to elucidate, in a unified way,  mechanisms and dynamics of localized objects out of equilibrium.

\section*{Dedication and Acknowledgements} 
It is our great pleasure to dedicate the present article to Professor Helmut R.\ Brand on the occasion of
his sixtieth birthday. Professor Brand has made outstanding efforts for the scientific interaction in the field of
non-linear and non-equilibrium physics between Japan and Germany for these thirty years. 

This work was supported by the JSPS Core-to-Core Program ``International research network for non-equi\-librium dynamics of soft matter" and  the Grant-in-Aid for the Global COE Program ``The Next Generation of Physics, Spun from Universality and Emergence" from the Ministry of Education, Culture, Sports, Science and Technology (MEXT) of Japan.
TO is supported by  a Grant-in-Aid for Scientific Research (C) from Japan Society for Promotion of Science. AMM acknowledges support from the Deutsche Forschungsgemeinschaft through project ``Nicht\-gleich\-ge\-wichts\-ph\"a\-no\-me\-ne in Weicher Materie/Soft Matter'' LO 418/15-1. 
MT is supported by Research Fellowships of the Japan Society for the Promotion of Science for Young Scientists.

\end{document}